\documentclass[final]{svjour3}
\usepackage{graphicx}
\usepackage{rotating}
\usepackage{amssymb}
\usepackage{siunitx}
\usepackage{array}
\newcolumntype{M}[1]{>{\centering\arraybackslash}m{#1}}
\DeclareSIUnit\sq{\ensuremath{\Box}}
\usepackage{mathptmx}
\usepackage[numbers,sort]{natbib}
\usepackage[colorlinks]{hyperref}

\makeatletter

\journalname{Journal of Low Temperature Physics}

\bibpunct{[}{]}{,}{n}{}{,}

\begin{document}

\newcommand{\hdblarrow}{H\makebox[0.9ex][l]{$\downdownarrows$}-}
\title{A kinetic inductance detectors array design for high background conditions at 150 GHz}

\author{Shibo Shu$^{1,\dagger}$ \and Jack Sayers$^{1}$  \and Peter Day$^2$ }

\institute{$^1$ California Institute of Technology, Pasadena, California 91125, USA\\
$^2$ Jet Propulsion Laboratory, California Institute of Technology, Pasadena, California 91109, USA\\
$^\dagger$\email{shiboshu@caltech.edu}}

\maketitle

\begin{abstract}
We present a design for an array of kinetic inductance detectors (KIDs) integrated with phased array antennas for imaging at 150 GHz under high background conditions. The microstrip geometry KID detectors are projected to achieve photon noise limited sensitivity with larger than 100~pW absorbed optical power. Both the microstrip KIDs and the antenna feed network make use of a low-loss amorphous silicon dielectric.  A new aspect of the antenna implementation is the use of a NbTiN microstrip feed network to facilitate impedance matching to the 50 Ohm antenna. The array has 256 pixels on a 6-inch wafer and each pixel has two polarizations with two Al KIDs. The KIDs are designed with a half wavelength microstrip transmission line with parallel plate capacitors at the two ends. The resonance frequency range is 400 to 800 MHz. The readout feedline is also implemented in microstrip and has an impedance transformer from 50 Ohm to 9 Ohm at its input and output.


\end{abstract}

\section{Introduction} 

Various designs have been proposed for applications from space projects~\cite{Griffin:2016SPACEKID,CORE2018,Glenn:2021GEP,OLIMPO2020flight,BLASTTNG2020} to ground-based large aperture telescopes~\cite{Sayers:2014a,Adam:2018a,Austermann:2018a, Tapia:2020MUSCAT} using kinetic inductance detectors~\cite{Day:2003a} at millimeter wavelength. Among these designs, the expected receiving power is usually smaller than 20~pW per detector, as the signal from sky is weak. Recently the need of using KIDs for high background conditions ($>$100~pW), such as terrestrial long-range imaging~\cite{Sayers:2020ONR} and security screening~\cite{Rowe:2016a}, is rising. In the long-range imaging case, the far-infrared radiation of targets is blocked by certain conditions such as fog, rain, and snow, so millimeter wave becomes a promising option. Their current development is based on a horn-coupled lumped-element KIDs design~\cite{Sayers:2020ONR,McCarrick:2018b}. The metal horn array is heavy, and has a different thermal expansion coefficient than silicon wafer. Also, it is hard to have both two-polarization and multi-band abilities using simple fabrication process, which is one of the main advantages of KIDs. Here we present an alternative design for this application at 150~GHz utilizing a low-loss amorphous silicon ($\alpha$-Si) dielectric~\cite{Golwala:poster}. $\alpha$-Si has a significant high dielectric constant, $\sim 10.3$ estimated at 1~K, which allows us to decrease the capacitor size necessarily. As a large inductor volume is needed for high background conditions, the kinetic inductance ratio is quite low and we need large capacitance to lower the resonance frequency and increase the sensitivity~\cite{Zmuidzinas:2012a}. Also, measurements show that $\alpha$-Si has a low two-level system (TLS) loss tangent of $\sim 10^{-5}$, and a low TLS noise~\cite{Golwala:poster}.

\section{Antenna and feed network design}

\begin{figure}
    \centering
    \includegraphics[width=0.95\textwidth]{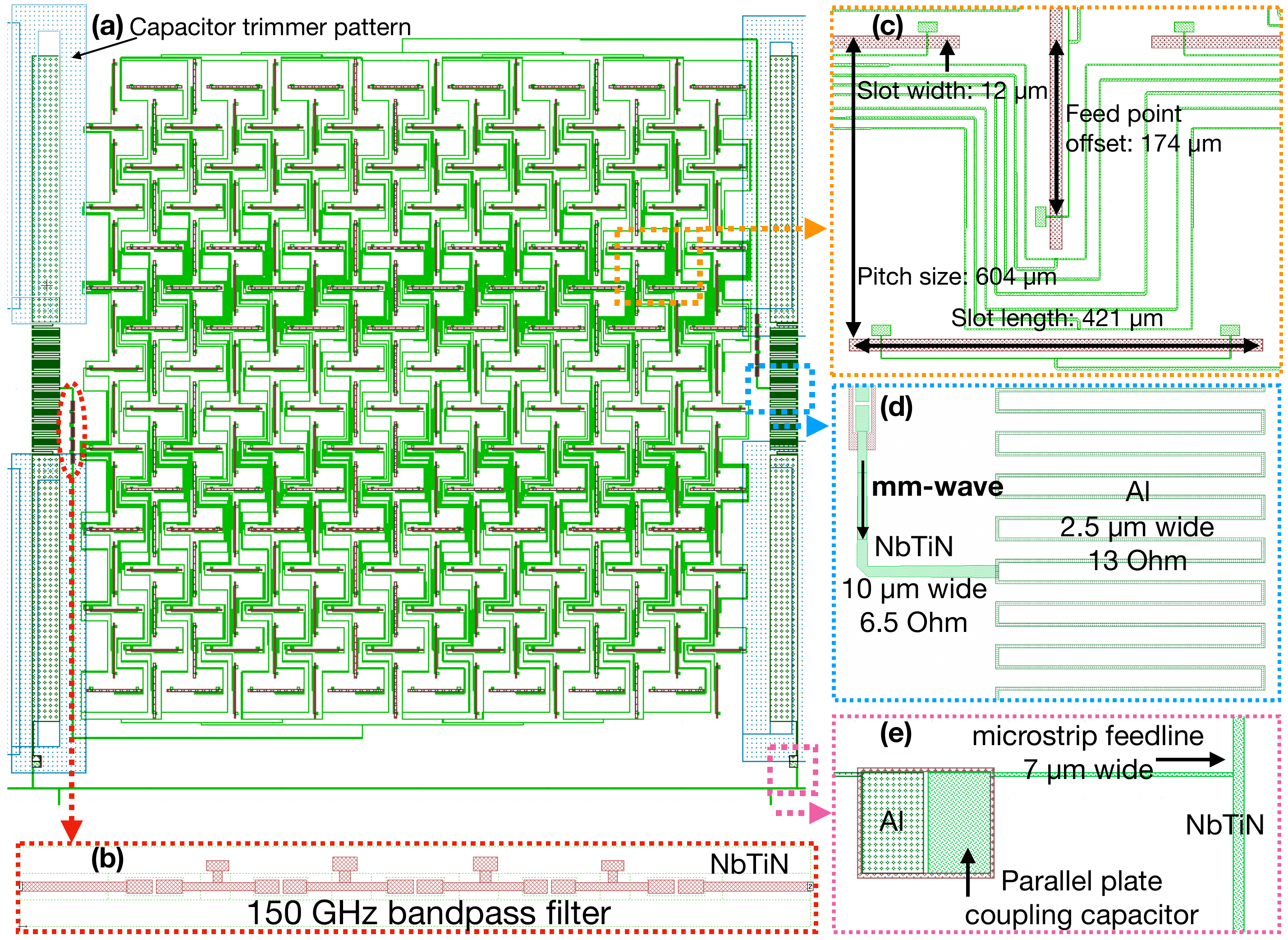}
    \caption{(a)Mask design of a single pixel. (b)Design of a 5-pole 150 GHz BPF using NbTiN as top metal. (c)Design of the slot antenna feed. (d)Millimeter wave is coupled from NbTiN microstrip line to the center of the Al microstrip resonator by impedance matching. (e)Parallel plate coupling capacitor. The ground plane is etched to be an island.}
    \label{fig:1}
\end{figure}

\begin{figure}
    \centering
    \includegraphics[width=0.95\textwidth]{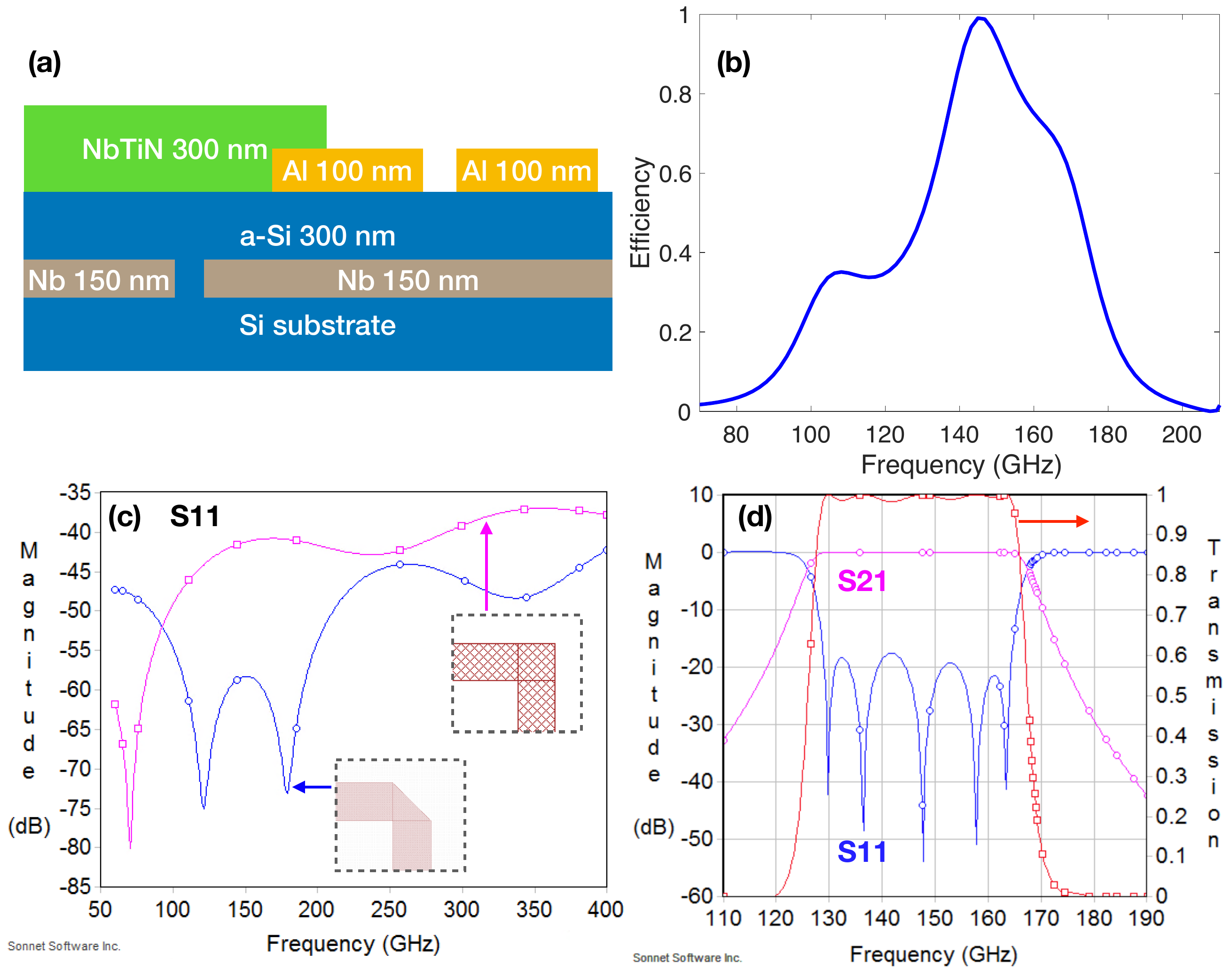}
    \caption{(a)Schematic drawing of the material stacking (not to scale). (b)Optical efficiency calculated from impedance matching at feed points. (c)The S11 of the mitered microstrip line bend. (d)Simulation of the 150 GHz BPF}
    \label{fig:2}
\end{figure}

The antenna design is based on slot antenna phased arrays~\cite{Zmuidzinas:1992a,Goldin:2004a,Ade:2014bicep2II}, shown in Fig.~\ref{fig:1}a. This antenna consists of $8\times 8$ slot antennas for each polarization and all slot antenna signals are added coherently. By placing the slot antennas in a certain way, two polarizations can be achieved. Each slot antenna has two microstrip feeds with details shown in Fig.~\ref{fig:1}c. A parallel plate capacitor is used at the feed to tune out the inductive impedance~\cite{Goldin:2004a}. The antenna transmission is calculated with an optimal feed impedance $47+15i$~Ohm using the method introduced in Ref.~\cite{Zmuidzinas:1992a}, and the result is shown in Fig.~\ref{fig:2}b. 

The high dielectric constant of $\alpha$-Si makes it hard to design a 47~Ohm microstrip line, for example the linewidth will be 0.6~$\mathrm{\mu m}$ using Nb, so here we choose NbTiN, which has a high kinetic inductance, as the microstrip line. The material stacking is shown in Fig.~\ref{fig:2}a. However, kinetic inductance strongly depends on the film thickness and a small thickness variation may result in a phase velocity difference inside the feed network of a 5~mm square pixel. After the trade-off, a 300~nm thick NbTiN with a kinetic inductance of 1~pH$/\ensuremath{\Box}$ is used and the linewidth for 47~Ohm microstrip line is 1.2~$\mathrm{\mu m}$. The $15i$~Ohm is realized by a parallel plate capacitor at the end of the microstrip line feed.

Mitered bends are used for all corners of the feed network to mitigate the reflection. The simulation in Fig.~\ref{fig:2}c shows that the in-band return loss of the bend is smaller than -55~dB. The crosstalk between two microstrip lines in parallel are optimized to be smaller than -30~dB with a center-to-center distance of 10~$\mathrm{\mu m}$ and a length of 1.5~mm, which are the closest distance and the longest length two parallel lines may have at the closest distance, respectively. In order to better restrict the band to the 2-mm atmospheric window, we have developed a 5-pole bandpass filter (BPF) design (Fig.~\ref{fig:1}b), which has been used in BICEP2~\cite{Ade:2014bicep2II}. The filter bandwidth is 126-167~GHz, shown in Fig.~\ref{fig:2}d.

The device will be fabricated on a high resistivity Si substrate with a thickness of 508~$\mathrm{\mu m}$. In calculation we assumed a 244~$\mathrm{\mu m}$-thick anti-reflection layer with a dielectric constant of 4.5, located on the back side of the wafer, and a backshort placed 517~$\mathrm{\mu m}$ away from the antennas on the front side of the wafer. This anti-reflection layer could be realized by a machined sub-wavelength structure on Si~\cite{Nitta:2017a,Defrance:2018arcoating}.

\section{Kinetic inductance detectors design}

\begin{figure}
    \centering
    \includegraphics[width=0.95\textwidth]{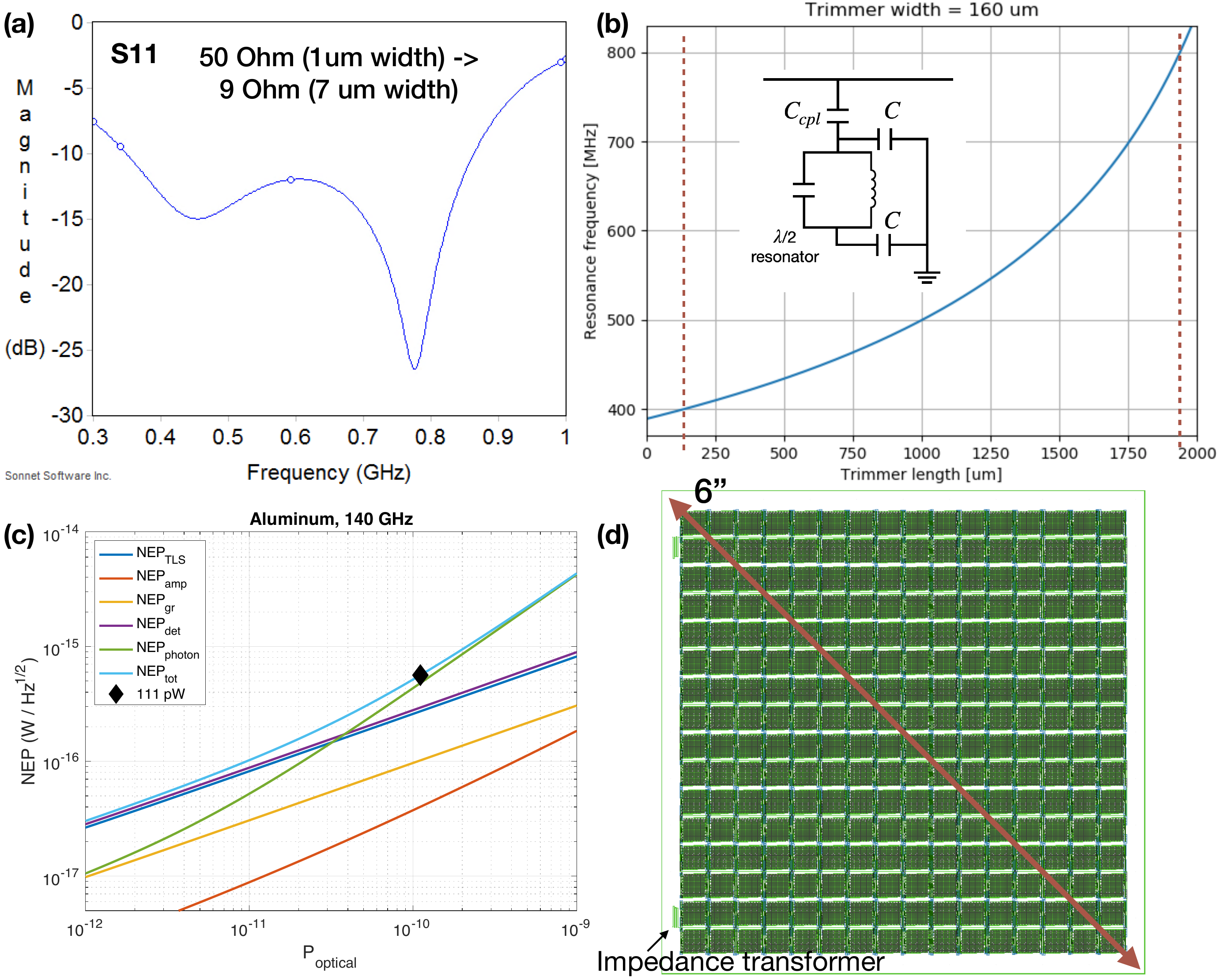}
    \caption{(a)Simulation of the stepped impedance transformer. (b)The relation of the resonance frequency and the trimmer length. (c)Calculation of all NEPs at different optical power. $\textrm{NEP}_{det}$ includes $\textrm{NEP}_{TLS}$, $\textrm{NEP}_{gr}$, and $\textrm{NEP}_{amp}$. A typical optical power is 111~pW, corresponding to 290~K. (d)The 256-pixel array design.}
    \label{fig:3}
\end{figure}

Al has a relative low kinetic inductance, so a thin film is needed to have a reasonable kinetic inductance ratio. For high background conditions, a large inductor volume is required to not saturate the detector. After calculation, we chose 100~nm thick Al, which has a sheet resistance of 0.069~Ohm$/\ensuremath{\Box}$ and a kinetic inductance of 0.11~pH$/\ensuremath{\Box}$. 

Our KID consists of a Al microstrip line as a half-wavelength resonator, and two parallel-plate capacitors for resonance frequency tuning (Fig.~\ref{fig:1}e). The inset in Fig.~\ref{fig:3}b shows the circuit model of a KID. All KIDs have the same resonator length of 13.2~mm and width of 2.5~$\mathrm{\mu m}$, giving a inductor volume of 3300~$\mathrm{\mu m^3}$. Given certain requirements like the readout bandwidth (400-800~MHz), multiplexing density (Q$>1\times 10^4$), and photon noise limited noise equivalent power (NEP), the design parameters are optimized at 200~mK, and the expected NEPs are shown in Fig.~\ref{fig:3}c.

Here we briefly present some parameters used in the NEP calculations, and the equations can be found in Ref.~\cite{Zmuidzinas:2012a}. The HEMT amplifier noise temperature is assumed to be 5~K. The readout power is set to be half the bifurcation power~\cite{Swenson:2012a}, given by
\begin{equation}
    P_r = \frac{1}{2} P_{bif} \approx \frac{1}{2} \frac{Q_c\omega_r E_*}{2Q_r^3},
\end{equation}
where $E_*$ is the scale of bifurcation and we assume it equals to the condensation energy $E_*=E_c=N_0 V \Delta^2/2$ for a rough estimation. $N_0=1.72\times 10^{10}$~eV/$\mu m^3$ is the single-spin density of states at the Fermi energy for Al, $V$ is the inductor volume, and $\Delta$ is the gap energy. The quasiparticles created by readout power is ignored. The internal quality factor $Q_i$ is calculated from quasiparticle density under certain optical power, and the coupling quality factor $Q_c$ is set to be the same with $Q_i$. The TLS noise $S_{TLS}=3\times 10^{-19}$$\mathrm{Hz}^{-1}$ is extracted from Ref.~\cite{Golwala:poster} at 200~mK and 100~Hz. At 111~pW, the photon NEP is $4.8\times 10^{-16}$~W/$\sqrt{\mathrm{Hz}}$, and the detector NEP is $2.9\times 10^{-16}$~W/$\sqrt{\mathrm{Hz}}$.

Two parallel plate capacitors are placed symmetrically at the two ends of each resonator to maintain a symmetric electric field inside the resonator. The area of the capacitor is calculated from the resonance frequency from 400~MHz to 800~MHz, shown in Fig.~\ref{fig:3}b. As we expect this design will be patterned using stepper lithography, which has limited area to cover the whole capacitor variations on a 6-inch wafer, we put 4 pixels (8 KIDs) into a group and a two-step exposure is planned. First the whole capacitor in positive mask will be patterned on a positive photoresist. Then we pattern the trimmers on capacitors where extra areas need to be trimmed. The trimmer patterns are designed to move along the long-grain direction of the capacitors for improving the tuning accuracy. The blank area between the KID capacitor and the coupling capacitor limits where trimmers can be placed, so 8 trimmer patterns with difference lengths are designed by separating the readout bandwidth into 8 groups. Also the coupling capacitors have 8 different values for the 8 frequency group, so we do not need to change the coupling capacitor area by sacrificing a bit sensitivity. 

Conventionally, the feedlines are designed with 50~Ohm transmission lines to match the impedance of coaxial cables. In our case, a 50~Ohm microstrip line has a width of 1~$\mathrm{\mu m}$, which is fragile across a 6-inch wafer. Using thinner NbTiN film would require an extra metal deposition. To make the feedline robust, we designed a 9~Ohm feedline with 7~$\mathrm{\mu m}$ wide NbTiN. A 6-section stepped impedance transformer~\cite{Matthaei:1966impedancetrans} is optimized for the readout bandwidth, with S11$<-12$ dB from 400~MHz to 800~MHz shown in Fig.~\ref{fig:3}a. The length of the transformer is 27~mm, which can be folded to have a small footprint (Fig.~\ref{fig:3}d). The coupling capacitors are calculated from this 9~Ohm feedline. 

The millimeter-wave signal is coupled to the resonator and get absorbed as it travels along the Al microstrip line. The coupling of millimeter-wave signal is realized by impedance matching from a 6.5~Ohm NbTiN microstrip line to the middle of the 13~Ohm Al microstrip line (Fig.~\ref{fig:1}d), where the resonance current has a maximum. At 150~GHz, the attenuation length in Al microstrip line is 0.5~mm, corresponding to 8.7~dB/mm. The half resonator length is 6.6~mm, so most of the power will be absorbed in the resonator and the coupling efficiency is mainly determined by the impedance mismatch at the T-junction.

\section{Conclusion}
We have shown a new kinetic inductance detectors array design for high background conditions at 150~GHz. The array is optimized to achieve photon noise limited sensitivity with larger than 100~pW absorbed optical power. The design utilize a low-loss amorphous silicon dielectric for the antenna and the KIDs. With certain modification, this design could also be used for low background conditions and other frequency ranges.

Data sharing not applicable to this article as no datasets were generated or analysed during the current study.

\bibliographystyle{JLTPv2}

\end{document}